# A TRANSFORMER-BASED DEEP Q LEARNING APPROACH FOR DYNAMIC LOAD BALANCING IN SOFTWARE-DEFINED NETWORKS

1st Evans Tetteh Owusu
*Telecommunication Engineering*
*Kwame Nkrumah University of Science and Technology*
Kumasi, Ghana
teowusu@st.knust.edu.gh

2nd Mr. Kwame Agyemang-Prempeh Agyekum
*Telecommunication Engineering*
*Kwame Nkrumah University of Science and Technology*
Kumasi, Ghana
kapagyekum.coe@knust.edu.gh

3rd Marina Benneh
*Telecommunication Engineering*
*Kwame Nkrumah University of Science and Technology*
Kumasi, Ghana
mbenneh@st.knust.edu.gh

4th Pius Ayorna
*Telecommunication Engineering*
*Kwame Nkrumah University of Science and Technology*
Kumasi, Ghana
payorna@st.knust.edu.g

5th Jusctice Owusu Agyemang
*Telecommunication Engineering*
*Kwame Nkrumah University of Science and Technology*
Kumasi, Ghana
jay@knust.edu.gh

6th George Nii Martey Colley
*Telecommunication Engineering*
*Kwame Nkrumah University of Science and Technology*
Kumasi, Ghana
gnmcolley@st.knust.edu.gh

7th James Dzisi Gadze
*Telecommunication Engineering*
*Kwame Nkrumah University of Science and Technology*
Kumasi, Ghana
jdgadze.coe@knust.edu.gh

*Abstract*— This study proposes a novel approach for dynamic load balancing in Software-Defined Networks (SDNs) using a Transformer-based Deep Q-Network (DQN). Traditional load balancing mechanisms, such as Round Robin (RR) and Weighted Round Robin (WRR), are static and often struggle to adapt to fluctuating traffic conditions, leading to inefficiencies in network performance. In contrast, SDNs offer centralized control and flexibility, providing an ideal platform for implementing machine learning-driven optimization strategies. The core of this research combines a Temporal Fusion Transformer (TFT) for accurate traffic prediction with a DQN model to perform real-time dynamic load balancing. The TFT model predicts future traffic loads, which the DQN uses as input, allowing it to make intelligent routing decisions that optimize throughput, minimize latency, and reduce packet loss. The proposed model was tested against RR and WRR in simulated environments with varying data rates, and the results demonstrate significant improvements in network performance. For the 500MB data rate, the DQN model achieved an average throughput of 0.275 compared to 0.202 and 0.205 for RR and WRR, respectively. Additionally, the DQN recorded lower average latency and packet loss. In the 1000MB simulation, the DQN model outperformed the traditional methods in throughput, latency, and packet loss, reinforcing its effectiveness in managing network loads dynamically. This research presents an important step towards enhancing network performance through the integration of machine learning models within SDNs, potentially paving the way for more adaptive, intelligent network management systems.

## I. INTRODUCTION

Software-Defined Networking (SDN) is transforming modern network management by decoupling the control plane from the data plane, enabling a more flexible, programmable, and efficient infrastructure for handling complex network environments (illustrated in figure 1). Traditionally, networking relied on hardware-based configurations where control logic was integrated into forwarding devices, limiting the flexibility and scalability of network operations. SDN addresses these limitations by separating decision-making from data forwarding processes, allowing centralized management and dynamic configuration of network resources. This centralized nature of SDN is crucial in today's networks, where traffic patterns can change rapidly, and efficient load balancing is critical for maintaining high performance[1], [2].

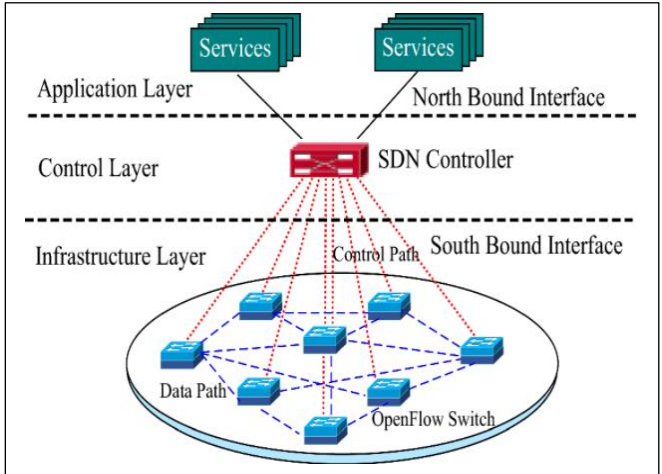

Fig. 1.        SDN Architecture[3].

However, load balancing in SDNs poses significant challenges, particularly due to the dynamic nature of network traffic. Traditional load balancing techniques, such as Round Robin (RR) and Weighted Round Robin (WRR), rely on static rules that do not adapt to fluctuating traffic conditions[4][5]. These methods often lead to suboptimal performance, with increased network latency, decreased throughput, and inefficient resource utilization. The need for adaptive load balancing solutions that can dynamically adjust to real-time traffic patterns is more pressing than ever, especially with the growth of data-driven applications and services[6].

To address these challenges, machine learning (ML) techniques have shown great promise in SDN environments[7][6]. ML models can process large amounts of network data, identify patterns, and make data-driven decisions that optimize traffic management. In this research, we propose a novel integration of the Temporal Fusion Transformer (TFT)[8] for traffic prediction and the Deep Q-Network (DQN) for dynamic load balancing. The TFT model is designed to capture long-range dependencies in time-series data, enabling accurate traffic forecasts, while the DQN is responsible for making real-time decisions that optimize traffic distribution across the network. By combining these models, the goal is to enhance network performance by reducing latency, improving throughput, and maximizing resource utilization.

The motivation behind this approach stems from the limitations of existing load balancing techniques in handling dynamic network environments. Traditional methods, which rely on static rules, struggle to adapt to changing traffic patterns, leading to congestion and inefficiencies[9]. Machine learning models, particularly those based on deep reinforcement learning, offer a solution to these challenges by learning from the network environment and making continuous adjustments. The inclusion of the TFT model further enhances the ability to predict future traffic loads, enabling the DQN to make more informed decisions about how to balance the load in real-time[10].

In SDN environments, the dynamic nature of network traffic often results in uneven load distribution, causing issues such as network latency and decreased throughput. Our approach leverages the capabilities of both transformers and deep reinforcement learning to create a more intelligent and adaptive load balancing solution. The proposed model can take into account both historical trends and forecasted future loads, enabling it to make more effective load balancing decisions compared to methods that only consider the network's current state.

The contributions of this research are fourfold. First, we introduce a novel combination of the Temporal Fusion Transformer (TFT) and Deep Q-Network (DQN) models for dynamic load balancing in SDN environments, an approach that has not been extensively explored in the literature. Second, we demonstrate through simulations that the TFT-DQN model significantly outperforms traditional methods, such as Round Robin (RR) and Weighted Round Robin (WRR), in terms of key performance metrics, including throughput, latency, and packet loss. Third, we conduct a comprehensive feature importance analysis, identifying critical variables such as Rx_Bandwidth_Utilization (Received Bandwidth Utilization), Tx_bitrate (Transmit Bitrate, Kbps), and Tx_packets (Transmit packets), which contribute most to traffic prediction. Lastly, the results from this study provide valuable insights for the future development of adaptive load balancing models, contributing to the advancement of machine learning applications in SDNs.

Given the rapid growth of network traffic driven by technologies like 5G, cloud computing, and the Internet of Things (IoT), the significance of this study is twofold[11]. First, it offers an intelligent, scalable solution for managing network resources efficiently in dynamic environments. Second, by demonstrating the effectiveness of integrating machine learning into SDN, this work opens the door for further research into real-time adaptive network management systems that can handle the increasing complexity of modern networks.

The proposed TFT-DQN model represents a significant step forward in dynamic load balancing for SDNs. Through accurate traffic forecasting and intelligent decision-making, this approach has the potential to enhance network performance, reduce congestion, and optimize resource utilization, paving the way for more adaptive and efficient network architectures in the future.

II. THEORETICAL BACKGROUND

A. *Software-Defined Networking (SDN)*

Software-Defined Networking (SDN) has transformed network management by decoupling the control plane from the data plane, allowing for centralized and dynamic control over network resources. This separation facilitates efficient optimization and reconfiguration, addressing challenges like scalability, flexibility, and security in modern networks, especially with the advent of 5G and beyond[12][2]. However, these evolving network demands necessitate more advanced load balancing mechanisms that can adapt to the complexities of real-time traffic conditions. SDN provides a centralized architecture that makes it an ideal platform for applying machine learning techniques, enabling dynamic load balancing and traffic optimization[6].

*B. Load Balancing in SDN*

Load balancing in SDN aims to distribute network traffic across multiple paths or servers, preventing congestion on any single component, shown in figure 2. While traditional load balancing techniques such as Round Robin (RR) and Weighted Round Robin (WRR) are effective in certain scenarios, they fail to adapt to dynamic and highly variable traffic patterns in complex network environments[4], [5], [9], [13]. To overcome these limitations, machine learning (ML) models are increasingly being adopted, offering real-time traffic prediction and dynamic load distribution, resulting in more intelligent and adaptive load balancing solutions[10]. The proposed approach leverages a Transformer-based Deep Q-Network (DQN) for dynamic load balancing, with the Temporal Fusion Transformer (TFT) forecasting traffic patterns that guide DQN in decision-making. This innovative combination improves SDN performance by capitalizing on the strength of transformers in capturing long-range temporal dependencies[8], [14].

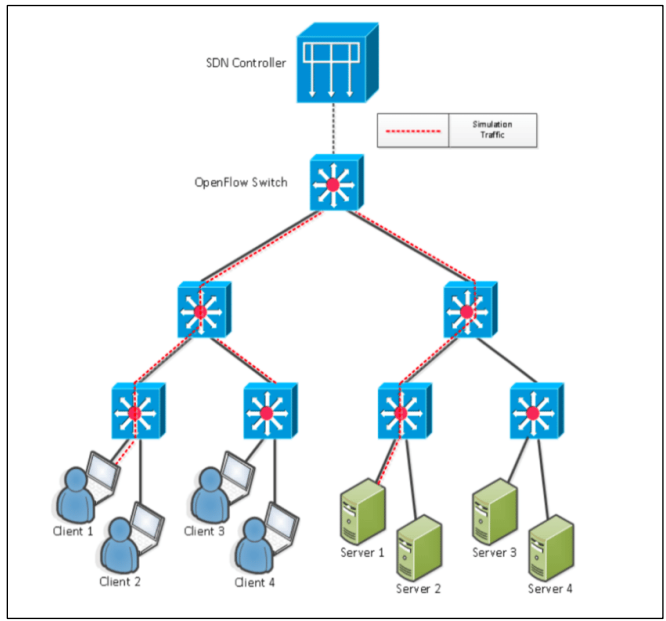

Fig. 2. Load Balancing in SDN[15]

### III. RELATED WORK

Several approaches have been proposed to address load balancing and traffic engineering in SDN and Software-Defined Wireless Networks (SDWN). These techniques fall into the following categories: Routing Optimization Algorithms, Clustering Algorithms, Supervised and Unsupervised Machine Learning, and Reinforcement Learning Techniques.

1) *Routing Optimization Algorithms*: Routing optimization in SDN remains a vital research area. [16] proposed a roadmap for traffic engineering in SDN-OpenFlow networks, enabling improved Quality of Service (QoS) through dynamic control of network flows[17]. Introduced in [18] is a load-balancing approach that adjusts traffic dynamically to reduce delays and enhance throughput. Expanding on these concepts, Filsfils and Previdi introduced Segment Routing, a forwarding architecture that simplifies packet routing by encoding path information in the packet headers[19]. Furthermore, Dobrijevic et al. applied Ant Colony Optimization (ACO) for flow routing in SDN, demonstrating the adaptability of ACO in dynamic environments[20].

2) *Clustering Algorithms:* Clustering algorithms have been applied to both SDN and SDWN environments to group similar traffic types and optimize traffic management based on QoS requirements. The authors utilized a semi-supervised machine learning approach for traffic classification, while Xiang proposed a clustering-based routing algorithm to optimize communication paths in wireless sensor networks[21][22].

3) *Supervised and Unsupervised Machine Learning:* Machine learning models, both supervised and unsupervised, have shown promise in tackling SDN load balancing issues. Srivastava and Pandey demonstrated the effectiveness of machine learning in optimizing traffic across multiple SDN controllers[23]. Kumar and Anand used supervised learning models to predict traffic and dynamically balance loads in real time[6]. Matlou and Abu-Mahfouzho incorporated AI techniques to improve traffic management in SDWN, while Kumar et al. presented an unsupervised learning method to detect failures and optimize network performance[10][6].

4) *Reinforcement Learning Techniques:* Reinforcement learning (RL) techniques are increasingly being adopted for traffic management in SDNs. Li et al. applied RL for SDN controller load balancing, where the controller learns optimal strategies for distributing traffic loads[24]. Tosounidis et al. employed deep Q-learning for traffic load balancing, improving network performance through a reward-based decision-making system[14]. Filali et al. and Almakdi et al. applied RL models in SDN-based 5G networks, demonstrating their effectiveness in managing dynamic network environments[25][11].

5) *Time-Series Forecasting and Transformers:* Recently, Transformer models have been used in time-series forecasting for SDNs to predict network traffic patterns. Wen et al. reviewed the application of Transformers in time-series forecasting, showing how attention mechanisms can capture long-term dependencies in network traffic data, thus enhancing load balancing and traffic prediction in SDNs[26][27].

## IV. SYSTEM MODEL

The system model presented in this work is structured into three main planes of the Software Defined Network (SDN) architecture: the Application Plane, Control Plane, and Data Plane. These planes work in concert to dynamically manage and optimize network traffic in real-time, utilizing both machine learning models and SDN controllers.

1) *Application Plane:* This layer houses two crucial components for traffic management:
   a) *Deep Q-Network (DQN) Load Balancer:* The DQN-based load balancer is responsible for selecting the optimal network paths, aiming to maximize throughput while minimizing latency and packet loss. The agent interacts with the controller, sending instructions for traffic flow management.
   b) *Temporal Fusion Transformer (TFT) Traffic Forecaster:* The TFT forecaster is tasked with predicting future network traffic patterns based on historical data. These predictions inform the DQN load balancer to make preemptive decisions on how traffic should be routed to maintain efficiency across the network.

   The application plane communicates with the control plane via the Northbound API (NB-API), transmitting the outputs from the machine learning models to the SDN controller.

2) *Control Plane*: In the control plane, the SDN Controller serves as the brain of the network, collecting real-time data from network devices and orchestrating how traffic is routed based on the decisions made by the application plane models. It communicates bi-directionally with the application plane through the NB-API and with the data plane via the Southbound API (SB-API).

3) *Data Plane:* The data plane consists of the networking devices, such as switches and routers, that forward the traffic between end-users and servers. The traffic patterns here are continuously monitored by the SDN controller, which informs the control plane about network conditions.

The figure 3 illustrates the flow of communication between the three planes. The SDN controller governs the network by issuing routing decisions to the switches, while also feeding back performance metrics like bandwidth utilization and packet loss to the DQN load balancer and TFT traffic forecaster, ensuring a closed-loop system that adapts to real-time traffic conditions.

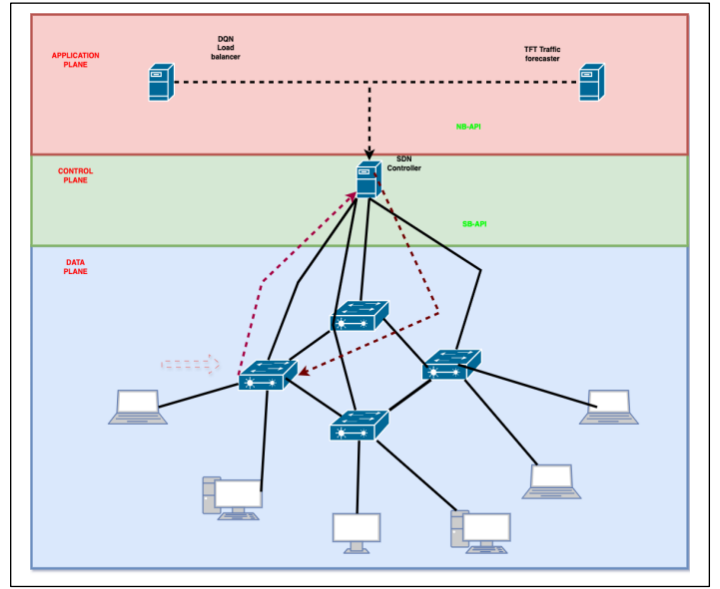

Fig. 3. The system Model

## V. METHODOLOGY

### A. Fat-Tree Topology

The fat-tree topology is a widely adopted network architecture in data canters, designed to overcome the limitations of traditional tree topologies, such as bottlenecks and single points of failure. It is structured as a multistage, hierarchical network consisting of three levels: core, aggregation, and edge layers illustrated in figure 4. The key feature of the fat-tree is that the bandwidth remains consistent across each layer, achieved by ensuring that each layer has an equal number of connections to the layer above it. This architecture allows for high scalability and uniform bandwidth distribution, making it ideal for environments with high traffic demands[28].

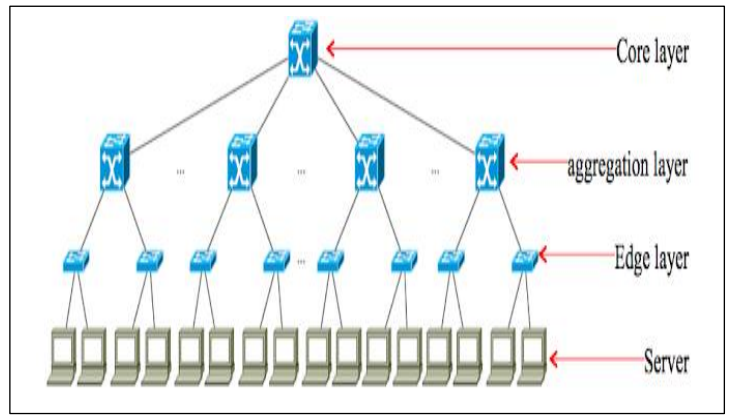

Fig. 4. Fat Tree Network Topology[29]

## B. Round Robin and Weighted Round Robin Algorithms

Round Robin (RR) and Weighted Round Robin (WRR) are two fundamental algorithms used for load balancing in network systems. The Round Robin algorithm distributes traffic evenly across servers or network paths by cycling through them in a fixed order, making it simple and easy to implement. However, RR does not account for the differing capabilities or current load of each server, which can lead to inefficiencies in scenarios with varying resource demands. Weighted Round Robin, an enhancement of RR, assigns different weights to each server or path based on their capacity or performance characteristics. This allows for a more balanced distribution of traffic, as servers with higher capacities can handle more requests. Despite its improvements over RR, WRR still has limitations, such as difficulty in adapting to real-time changes in server performance or load, and the complexity of determining appropriate weights, which can impact its effectiveness in dynamic environments..

## C. Time Series Traffic

Time series traffic refers to the modeling and analysis of network traffic data that is collected and observed sequentially over time. In network environments, traffic data typically exhibits temporal patterns, where the traffic load at any given moment is influenced by past events and trends. These patterns can include hourly or daily cycles, periodic spikes, and long-term trends that need to be captured accurately for effective network management. Time series analysis in network traffic is crucial for understanding traffic dynamics, detecting anomalies, and making informed decisions about resource allocation and load balancing. Techniques such as autoregressive models, moving averages, and more recently, advanced deep learning models like Temporal Fusion Transformers (TFT), are used to predict future traffic loads based on historical data. Accurately forecasting time series traffic helps in anticipating congestion, optimizing routing paths, and improving overall network performance by enabling proactive management strategies[30].

## D. Multi-Step Traffic Prediction

Multi-step traffic prediction involves forecasting network traffic over multiple future time steps, rather than predicting a single future point. This approach is particularly valuable in scenarios where network management requires planning over extended periods, such as in dynamic load balancing, capacity planning, and congestion control. In multi-step prediction, the accuracy of the forecast becomes progressively more challenging as the prediction horizon extends, due to the compounding of errors and the increasing uncertainty over time. To address this, models such as Recurrent Neural Networks (RNNs), Long Short-Term Memory (LSTM) networks, and Temporal Fusion Transformers (TFT) are employed to capture long-term dependencies and trends in the data. Multi-step predictions can be done using either direct methods, where separate models are trained for each prediction step, or recursive methods, where a single model predicts one step ahead and the prediction is fed back into the model for subsequent steps. Effective multi-step traffic prediction allows network administrators to better anticipate and manage future traffic conditions, leading to more efficient and resilient network operations[31], [32].

## E. Simulation Setup

In our simulation setup, we employed two virtual environments hosted on an Ubuntu machine, with one environment dedicated to running the Ryu controller and the other to Mininet. Both virtual devices were configured to support OpenFlow protocols, which facilitated seamless communication and control within the Software-Defined Networking (SDN) architecture. The Ryu controller was chosen for its versatility and comprehensive support for OpenFlow, enabling us to effectively manage and monitor network traffic, as well as to collect detailed statistics.

The network topology implemented in Mininet was based on a fat-tree architecture, featuring 16 hosts connected through a hierarchical structure of core, aggregation, and edge switches. The topology script initiated the network, followed by the configuration of queue rates on all switches to simulate varying network conditions. This allowed us to emulate real-world traffic scenarios, essential for evaluating the dynamic load balancing capabilities of our approach. The simulation included the generation of diverse traffic types—TCP, UDP, HTTP, DNS, and ICMP—across the hosts to test the network's response under different protocols and data loads. Each host was set up to serve an HTML file, and traffic was generated using tools like iperf, wget, dig, and ping.

---

**Module 1:** Importing libraries

1. from mininet.net import Mininet
2. from mininet.node import RemoteController, OVSKernelSwitch, Host
3. from mininet.cli import CLI
4. from mininet.log import setLogLevel, info
5. from time import sleep
6. import csv
7. import os

---

Mininet: represents the core Mininet class for creating and managing networks.

RemoteController: this allows adding remote controllers to the network.

OVSKernelSwitch: this helps to specify the switch type to use. In this case, OVS was used.

Host: this allows addition of hosts to the network.

Info: this creates a logging function to output messages to the console.

CLI: this enables the initialization of an interactive command-line interface to interact with the network.

**Module 2:** Custom Topology Initiation

1. def treeTopo( ) :
2.     net = Mininet(controller=RemoteController, switch=OVSSwitch, link=TCLink)
3.     net.addController('c0', controller=RemoteController, ip='127.0.0.1', port=6633)

To gather performance metrics, the Ryu controller was equipped with a script to collect flow, port, and queue statistics at regular intervals of 10 seconds. These statistics were crucial for analyzing the efficiency of load balancing and were saved in a CSV file for further analysis. The simulation was designed to run for a total of 12 hours, ensuring that sufficient data was collected to assess the network's behaviour over an extended period, under various load conditions. This setup allowed us to rigorously test and validate our proposed load balancing algorithms within a controlled, yet realistic, network environment.

**Module 3:** Importing libraries for ryu

1. import csv
2. import datetime
3. import time
4. from ryu.app import simple_switch_13
5. from ryu.controller import ofp_event
6. from ryu.controller.handler import MAIN_DISPATCHER, DEAD_DISPATCHER
7. from ryu.controller.handler import set_ev_cls
8. from ryu.ofproto import ofproto_v1_3
9. from ryu.lib import hub
10. from operator import attrgetter

The parameters used in this simulation are summarized in Table I.

TABLE I: Simulation Parameters

| Parameter | Description | Value |
|---|---|---|
| **Network Setup** | | |
| net | Network creation using Mininet | Mininet(controller= RemoterController, switch=OVSSwitch, link=TClink) |
| controller | Controller added to the network | RemoteController |
| controller IP | IP address of the remote controller | 127.0.0.1 |
| controller port | Port address of the remote controller | 6633 |
| **Queue Configuration** | | |
| min_rate | Minimun rate for queues | 5,000,000 (switches s7, s8, s9, s10); 1,000,000 (other switches) |
| max_rate | Maximum rate for queues | 10,000,000 (switches s7, s8, s9, s10); 5,000,000 (other switches) |
| **Traffic Simulation** | | |
| **TCP Traffic** | Protocol used for TCP Traffic | iperf |
| -p 5001 | Port for TCP Traffic | 5001 |
| -t 60 | Duration of the TCP traffic test | 60 seconds |
| -b 500M | Bandwidth for TCP traffic | 500M ( 500Mbps ) |
| **UDP Traffic** | Protocol used for UDP Traffic | iperf |
| -u | UDP mode | Enabled |
| -p 5002 | Port for UDP Traffic | 5002 |
| -t 60 | Duration of the UDP traffic test | 60 seconds |
| -b 500M | Bandwidth for UDP traffic | 500M ( 500Mbps ) |
| HTTP Traffic | Protocol used for HTTP Traffic | HTTP |
| Port for HTTP traffic | Port number used for HTTP Traffic | 80 |
| HTTP Server | Command to start HTTP server | python -m SimpleHTTPServer 80 & |
| HTTP Client | Command to request the HTTP page | wget -O /dev/null https://<host-IP>/index/html |
| DNS Traffic | Protocol used for DNS Traffic | DNS |
| DNS Query | Command to simulate DNS server | dig @<host-IP>www.example.com |
| ICMP Traffic | Protocol used for ICMP Traffic | ICMP |
| ICMP Ping | Command to simulate ICMP traffic | ping -c 100 <host-IP> & |
| Number of ICMP packets sent | Number of ping packets sents | 100 |

*F. Dataset Preparation and Description*

The dataset used in this study was meticulously constructed by integrating flow statistics, port statistics, and queue statistics, resulting in a comprehensive dataset comprising 1,185,808 data points. To ensure the quality and integrity of the data, extensive feature engineering was performed, including the imputation of missing values and the handling of outliers by replacing them with the average of the surrounding sequence data. Numerical features were normalized to facilitate consistent data analysis.

*G. TFT Modelling*

The Temporal Fusion Transformer (TFT) model, used in this work for network traffic forecasting, is a sophisticated architecture designed to capture complex temporal dependencies and variable interactions in time series data. The model's architecture comprises several key components, each serving a distinct function in processing the input features and generating accurate predictions. Below is a detailed explanation of the various layers and mechanisms used in the TFT architecture, including their mathematical formulations, as referenced from the original paper "Temporal Fusion Transformers for Interpretable Multi-horizon Time Series Forecasting" by Bryan Lim et al [8].

Figure 5 represents the architecture of the Temporal Fusion Transformer (TFT) used for traffic forecasting in the proposed simulation model. This model utilizes a series of input features, including historical traffic data, to predict future network load, which aids in optimizing load balancing decisions within the Software-Defined Network (SDN).

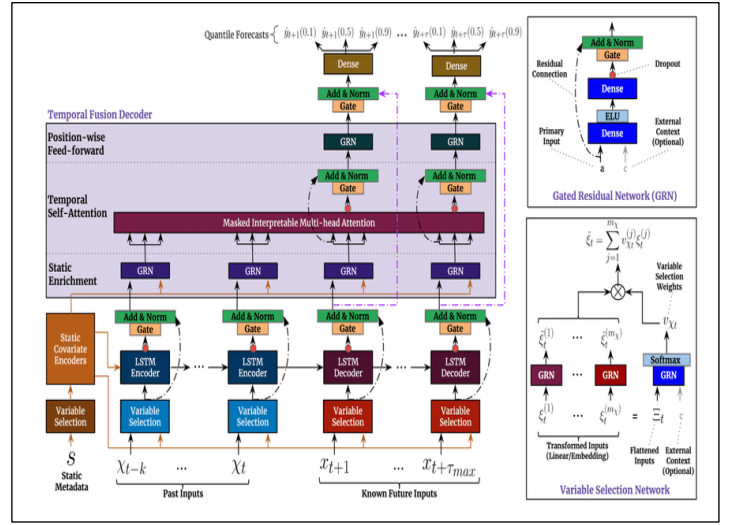

Fig. 5. Architecture of the Temporal Fusion Transformer (TFT)

1) *Gating Mechanism:* The gating mechanism in the TFT is a crucial component that helps control the flow of information through the network. It allows the model to adaptively select the most relevant features at each time step, thereby improving the model's robustness and interpretability. The gating mechanism can be mathematically represented as:

$$Gate(x) = \sigma(W_g \cdot x + b_g) \odot x$$

where:
  i. $W_g$ and $b_g$ are learnable parameters (weights and biases).
  ii. σ is the sigmoid activation function, which scales the input values between 0 and 1.
  iii. ⊙ denotes element-wise multiplication.

This mechanism ensures that irrelevant features are suppressed while allowing important features to pass through, which enhances the model's ability to focus on meaningful patterns in the data.

2) *Variable Selection Network (VSN):* The Variable Selection Network is designed to dynamically select the most important variables from both static and time-varying covariates. The VSN applies a soft attention mechanism to each feature, which can be expressed as:

$$a_i = softmax(W_{vs} \cdot x_i)$$

where:
i. $W_{vs}$ is the weight matrix specific to the variable selection layer.
ii. $x_i$ represents the input features.

The softmax function normalizes the attention scores $\alpha_i$ across all features, ensuring that the model focuses on the most relevant variables at each time step. This selective attention allows the TFT to efficiently handle large sets of input features, prioritizing those that contribute most to the prediction task.

3) *Static Covariant Encoders:* Static covariates, such as Link_id and Eth_dst in the context of this work, are features that remain constant throughout the time series. These features are encoded using static covariate encoders, which allow the model to incorporate long-term characteristics that do not vary over time. The static covariates are processed through a fully connected layer to produce embeddings, which are then concatenated with the dynamic features to provide additional context to the model.

4) *Interpretable Multi-head Attention:* The TFT leverages a multi-head attention mechanism to capture complex interactions between different time steps. This attention mechanism can be mathematically described as:

$$Attention(Q, K, V) = softmax\left(\frac{QK^T}{\sqrt{d_k}}\right)V$$

where:
Q (query), K (key), and V (value) are projections of the input features.
$d_k$ is the dimensionality of the key vectors.

The multi-head attention mechanism allows the model to focus on different aspects of the input sequence by applying attention in parallel across multiple heads. This results in a richer representation of the input data, enabling the model to capture both short-term and long-term dependencies in the time series.

5) *Temporal Processing Layer:* The TFT incorporates temporal processing layers, such as the GRU (Gated Recurrent Unit), to model sequential dependencies within the time series. The GRU layer processes the input sequence iteratively, capturing temporal patterns that are crucial for accurate forecasting. The GRU update equations are:

$$z_t = \sigma(W_z \cdot [h_{t-1}, x_t])$$
$$r_t = \sigma(W_r \cdot [h_{t-1}, x_t])$$

$$\hat{h}_t = tanh\left(W_h \cdot [r_t \odot h_{t-1}, x_t]\right)$$

$$h_t = (1 - z_t) \odot h_t - 1 + z_t \odot \hat{h}_t$$

These equations allow the GRU to efficiently capture temporal dependencies while mitigating the vanishing gradient problem, which is common in deep recurrent networks.

6) *Quantile Output:* The TFT model outputs predictions in the form of quantiles, which provide a probabilistic range of possible outcomes. This is particularly useful for forecasting tasks where uncertainty is inherent. The quantile loss function used in the TFT can be expressed as:

$$Quantile\ Loss = max(\tau \cdot (y - \hat{y}), (1 - \tau) \cdot (\hat{y} - y))$$

where:
i. y is the actual value.
ii. ŷ is the predicted value.
iii. τ is the quantile being predicted.

This loss function penalizes underestimation and overestimation differently, depending on the quantile, which allows the model to produce more informative forecasts by capturing the distribution of possible future values.

## VI. IMPLEMENTATION AND TRAINING

The implementation of the TFT in this work follows a structured approach, where the dataset is divided into training and validation sets based on a time index. The model is trained using a PyTorch Lightning framework, which handles the training loop, logging, and gradient clipping. The learning rate is optimized using a learning rate finder, which systematically explores different learning rates and suggests the best one based on observed loss trends. The suggested learning rate is then used to train the model, ensuring that the optimization process is both efficient and effective, leading to better convergence and improved forecasting performance. After several rounds of training and hyperparameter tuning, the most important hyperparameters of the retained TFT model are summarized in Table II.

Table II: Most Important Hyperparameters

| Hyperparameter | Value |
|---|---|
| Batch size | 128 |
| Number of workers | 2 |
| Maximum prediction length | 12 |
| Maximum encoding length | 24 |
| Hidden size | 8 |
| Attention head size | 1 |
| Dropout | 0.1 |
| Hidden Continuous size | 8 |
| Loss | Quantile loss |
| Log interval | 2 |
| Suggested Learning rate | 6.6069345e-5 |

### A. LSTM Modeling

Long Short-Term Memory (LSTM) networks are a specialized form of Recurrent Neural Networks (RNNs), designed to address the issue of long-term dependencies. Traditional RNNs suffer from vanishing and exploding gradient problems when learning long-range patterns due to their simple structure, making it difficult for them to retain information across long sequences. LSTMs solve this by introducing memory cells that can store information over long durations and gates that control the flow of information into and out of these cells[33].

The key components of an LSTM are:
1) *Cell State:* This is the core of the LSTM unit, responsible for maintaining long-term dependencies. The cell state allows information to flow unchanged through the entire sequence unless modified by the gates.
2) *Forget Gate:* This gate decides what information to discard from the cell state. It takes the current input and the previous hidden state and outputs a value between 0 and 1, where 0 means "completely forget" and 1 means "completely retain."

$$f_t = \sigma(W_f \cdot [h_{t-1}, x_t] + b_f)$$

Here, σ is the sigmoid activation function, $h_{t-1}$ is the previous hidden state, $x_t$ is the current input, and $W_f$ and $b_f$ are learnable parameters.

3) *Input Gate*: This gate controls what new information to store in the cell state. It has two parts: the sigmoid layer that decides which values to update and a tanh layer that creates a vector of new candidate values to add to the state.

$$i_t = \sigma(W_i \cdot [h_{t-1}, x_t] + b_i)$$

$$C_t = tanh(W_C \cdot [h_{t-1}, x_t] + b_c)$$

4) *Update of Cell State*: The cell state is updated by combining the forget gate's decision and the new candidate values. Part of the old cell state is retained (as determined by the forget gate), and new information (as determined by the input gate) is added.

$$C_t = f_t * C_{t-1} + i_t * Ć$$

5) *Output Gate:* This gate decides what part of the cell state to output. The output is based on the current input and the previous hidden state, passed through a sigmoid layer. The filtered cell state is then passed through a tanh function to produce the new hidden state.

$$o_t = \sigma(W_o \cdot [h_{t-1}, x_t] + b_o)$$

$$h_t = o_t * tanh(C_t)$$

LSTMs are crucial in tasks involving sequential data, such as time-series forecasting, which aligns with the goal of dynamic load balancing in SDN. By retaining and learning long-term dependencies, LSTMs can capture the trends and patterns in network traffic data over time, allowing for better decision-making in load balancing. The ability to remember past traffic fluctuations for long periods enhances the model's ability to predict future network states.

### B. Implementation of LSTM Training

The LSTM implementation follows the same dataset and learning rate as the Temporal Fusion Transformer (TFT) model. The dataset includes time-indexed traffic data, and the features have been preprocessed, including normalization and feature selection. The training involves minimizing a loss function such as Mean Squared Error (MSE) over the predictions and actual values in the dataset.

The LSTM model is trained using the following parameters:
  a) *Dataset:* Same as used for TFT, containing traffic load, latency, packet loss, and throughput metrics.
  b) *Learning Rate:* Same learning rate as the TFT model.
  c) *Optimizer:* Adam optimizer for efficient training.
  d) *Batch Size:* Consistent with TFT for fair comparison.

e) *Epochs:* The number of epochs 64, determined to ensure convergence without overfitting.

## C. DQN Modeling

To balance the load across the network dynamically in real-time, a Deep Q-Network (DQN) was employed. This section begins by providing the theoretical background on reinforcement learning, Q-learning, convolutional neural networks (CNNs), and the architecture of DQN, followed by the implementation and training of the model[34].

1) *Reinforcement Learning (RL):* Reinforcement learning (RL) is a branch of machine learning where an agent learns to make decisions by interacting with an environment. The agent takes actions in various states of the environment and receives feedback in the form of rewards. The goal of RL is to learn a policy that maximizes the cumulative reward over time, also called the return. Mathematically, RL problems are often modeled as Markov Decision Processes (MDPs), where the state transitions depend only on the current state and action, not on past states (the Markov property). Formally, an MDP is defined by the tuple (S, A, P, R, γ),
where:
a) S is the set of states,
b) A is the set of actions,
c) P(s'| s, a) is the state transition probability from state s to state s' given action a,
d) R(s, a) is the reward function that gives the immediate reward for taking action α in state s,
e) γ is the discount factor that determines how much future rewards are weighted compared to immediate rewards.

The agent's objective is to maximize the cumulative discounted reward, which can be represented as:

$$G_t = R_{t+1} + \gamma R_{t+2} + \gamma^2 R_{t+3} + \cdots = \sum \gamma^k R_{t+k+1}$$

Where $G_t$ is the return at time t.

2) *Q-Learning:* Q-learning in figure 6, is a popular model-free RL algorithm that seeks to learn an optimal action-selection policy. It aims to estimate the action-value function, also known as the Q-function, which represents the expected return of taking action aa in state ss and following the optimal policy thereafter. The Q-function is given by:
$$Q(s,a) = E[G_t | s_t = s, a_t = a]$$

The Q-learning algorithm updates its Q-values iteratively using the Bellman equation:

$$Q(s,a) \leftarrow Q(s,a) + \alpha[r + \gamma maxa'Q(s',a') - Q(s,a)]$$

where α is the learning rate, and γ is the discount factor. Over time, the Q-values converge to the optimal Q-function, which can then be used to derive the optimal policy by selecting the action with the highest Q-value in each state.

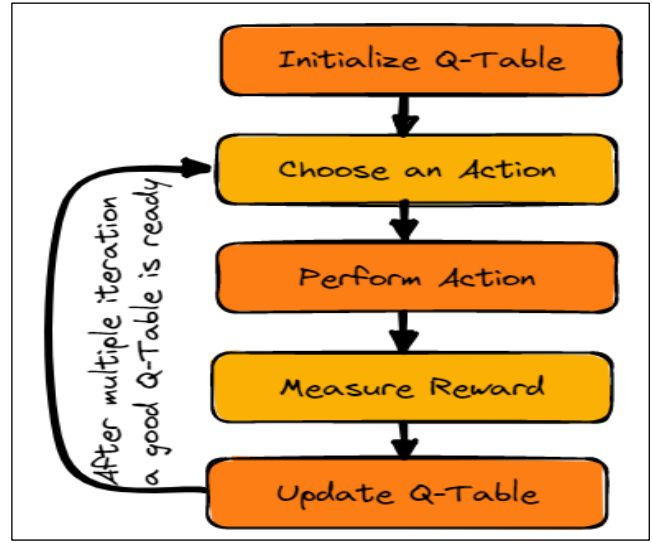

Fig. 6. Q Learning process[35].

3) *Convolutional Neural Networks (CNNS):* CNNs are a type of neural network architecture commonly used for processing grid-like data, such as images. In the context of RL and DQN, CNNs are utilized to process large state spaces, such as images or high-dimensional inputs. The convolution operation in CNNs allows the network to learn spatial hierarchies by applying filters to local patches of the input data. CNNs consist of multiple layers, including convolutional layers, pooling layers, and fully connected layers, which help capture both low-level and high-level features of the data.

4) *Architecture of DQN:* DQN combines Q-learning with deep learning, specifically using a deep neural network (figure 7) to approximate the Q-function. In traditional Q-learning, maintaining a Q-table is infeasible for large state-action spaces, but DQN overcomes this limitation by using a neural network to predict Q-values.
The DQN architecture typically consists of:
a) *Input Layer:* The input to the network is the state of the environment, which can be high-dimensional, such as a sequence of network statistics (e.g., latency, throughput).
b) *Convolutional Layers:* These layers are used when the input is an image or grid-like data. For general data, fully connected layers can be employed instead.
c) *Fully Connected Layers:* After feature extraction via convolutional layers, the output is passed through fully connected layers to learn the relationship between the input features and Q-values.

d) *Output Layer:* The output layer contains one node for each possible action, representing the Q-values for each action given the current state.

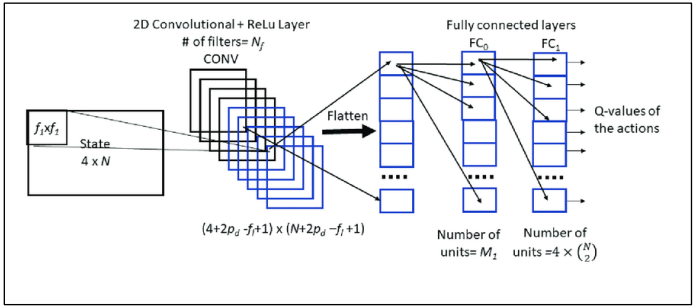

Fig. 7. Architecture of DQN[36].

One critical feature of DQN is the use of experience replay, where transitions (s, a, r, s′) are stored in a replay buffer and sampled randomly during training. This helps break the correlation between consecutive training samples and improves training stability. Additionally, DQN uses a target network, which is a copy of the Q-network that is updated less frequently, providing more stable Q-value updates.

The loss function for training the DQN is based on minimizing the temporal difference (TD) error:

$$L(\theta) = E_{(s,a,r,s')}[(r + \gamma \max a' \, Q_{target}(s',a',;\theta^-) - Q(s,a;\theta))^2]$$

where $\theta$ are the parameters of the Q-network, and $\theta^-$ are the parameters of the target network.

In the next section, the implementation and training of the DQN model will be discussed, detailing how the architecture was applied to achieve dynamic load balancing in the Software-Defined Network (SDN) environment.

### D. Implementation and Training of DQN

In this project, a custom gym environment was created to simulate the dynamics of a software-defined network (SDN) for load balancing, with a focus on maximizing throughput while minimizing latency and packet loss. The environment is based on a fat-tree topology, which includes 16 hosts, 2 core switches, 4 aggregation switches, and 8 edge switches. The SDN links in this environment carry specific values for throughput, latency, packet loss, and forecasted traffic. The agent's goal is to intelligently choose the best path (link) with the least forecasted traffic to route incoming network traffic. This process is formulated as a reinforcement learning problem where an agent learns through interacting with the environment by choosing actions and receiving feedback in the form of rewards.

1) *Network Simulation Environment:* The environment was designed using the OpenAI Gym interface, allowing easy integration with reinforcement learning algorithms. A CSV file containing the network's state data was used to load various metrics, including throughput, latency, packet loss, and forecasted traffic, for each link in the network. The environment is structured as follows:

a) *State Space:* Each state represents a snapshot of the network's throughput, latency, packet loss, and forecasted traffic for each link. The state space consists of a 2D matrix where each row represents a link, and the columns represent the metrics associated with that link. The state is normalized to bring the metrics into a uniform range between 0 and 1.

b) *Action Space:* The action space is discrete, where each action corresponds to selecting a specific link for routing traffic. There are as many actions as there are links in the network.

c) *Reward Function:* The reward function encourages the agent to maximize throughput while minimizing latency and packet loss. The reward is calculated as:

$$Reward = \alpha \sum throughput - \beta \sum latency - \gamma \sum packetLoss$$

This reward structure incentivizes the agent to select links that balance high throughput with low latency and packet loss.

2) *Network Topology and Link Generation:* The fat-tree topology was generated by defining connections between core switches, aggregation switches, edge switches, and hosts. The environment uses NetworkX to create and visualize the network as a directed graph, where:

a) Core switches are connected to all aggregation switches.
b) Aggregation switches are connected to specific edge switches.
c) Edge switches are connected to two hosts each.

Each link in this topology has associated metrics for throughput, latency, packet loss, and forecasted traffic, which are updated with each episode of the environment as shown in figure 8.

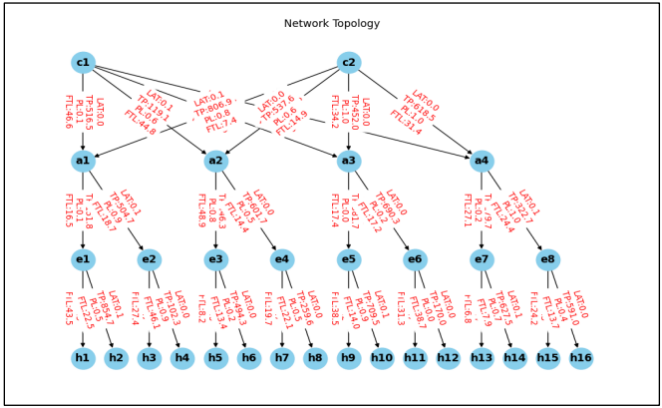

Fig. 8. A State of the Network Environment

3) *Agent Training with Deep Q Network (DQN):* The agent was trained using a Deep Q-Network (DQN) model. The DQN leverages deep learning techniques to estimate the Q-values for each possible action (link selection) given the current state of the network. The architecture of the DQN consists of fully connected layers that take the flattened state as input and output a Q-value for each action.

The training process is illustrated in figure 9.

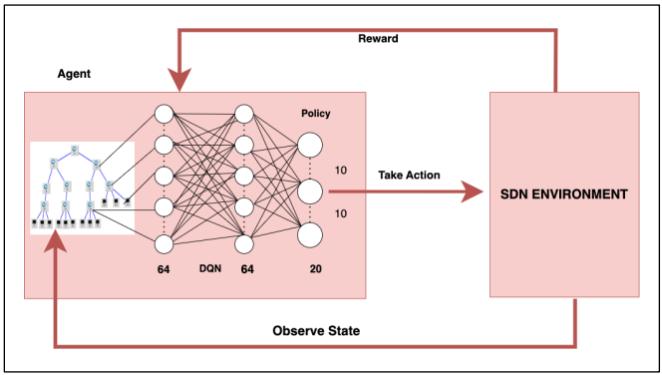

Fig. 9. DQN Training

a) *Network Architecture:* The DQN consists of:
   i. A fully connected layer with 128 units and a ReLU activation function.
   ii. A dropout layer to prevent overfitting.
   iii. Another fully connected layer with 128 units and a ReLU activation function, followed by a second dropout layer.
   iv. A final output layer that maps the hidden representation to Q-values for each action.

b) *Replay Buffer:* To improve sample efficiency and mitigate correlation in consecutive experiences, a replay buffer was implemented. The replay buffer in figure 10 stores past experiences (state, action, reward, next state, done) and samples mini-batches for training the DQN. This helps stabilize learning by breaking the temporal correlations between consecutive experiences.

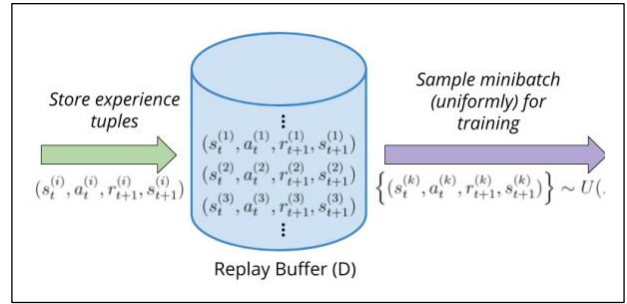

Fig. 10. Replay Buffer

c) *Training Process:* During training, the agent follows an epsilon-greedy policy, where it selects a random action with probability ε (exploration) and the action with the highest Q-value with probability 1−ε (exploitation). As training progresses, ε decays from 1.0 to 0.01, reducing exploration and favoring exploitation.

At each time step:
   i. The agent observes the current state (network metrics).
   ii. The agent selects an action (link) based on the Q-values.
   iii. The environment transitions to the next state and returns a reward.
   iv. The experience is stored in the replay buffer.

If enough experiences have been gathered, the agent samples a mini-batch from the buffer and performs gradient descent on the loss between the predicted Q-values and the target Q-values. The target Q-values are computed using the Bellman equation:

$$Q(s, a) = r + \gamma \max a' \, Q'(s', a')$$

where $Q'(s', a')$ are the Q-values predicted by a target network. The DQN weights are updated to minimize the mean squared error between the predicted Q-values and the target values.

d) *Target Network:* A target network was used to stabilize training. The target network is a copy of the DQN that is updated less frequently, providing more stable targets for the Q-learning updates.

The hyperparameter used in the DQN training is structured in table III.

Table III: Hyperparameters used in DQN training

| Hyperparameter | Values |
|---|---|
| Number of episodes | 1000 |
| Replay buffer size | 10,000 |
| Batch size | 64 |
| Learning rate | 0.003 |
| Discount factor($\gamma$) | 0.99 |
| Initial exploration rate($\varepsilon$) | 1.0 |
| Final exploration rate | 0.01 |
| Exploration decay | 0.995 |

Throughout the training process, metrics such as throughput, latency, packet loss, and reward were tracked. The DQN gradually learned to select links with lower forecasted traffic, achieving better load balancing and network performance by optimizing the overall throughput and minimizing the latency and packet loss.

The best-performing model was selected based on the cumulative reward and improvements in throughput, latency, and packet loss over time. The agent's actions increasingly aligned with network optimization objectives as training progressed.

Algorithm 1 was used to train the Deep Q Network Learning Model.

**Algorithm 1:** Deep Q Learning Training

1. Initialize Q(s, a) and weights with random normal distribution and Xavier initializer.
2. For each episode:
   Reset the environment and get an initial state s.
   While not done:
      With probability $\epsilon$, select random action a, otherwise select max(Q(s, a)).
      Execute action a in the environment, obtain reward r and next state s'.
      Store transition (s, a, r, s', done) in the replay buffer.
      If episode length > 1000:
         Sample a random mini-batch of transitions from the replay buffer.
         For each transition in the mini-batch:
            Calculate the target Q = r + $\gamma$ (Q(s', a)) if the episode not done,
      else
         target $Q = r$.
         Calculate $cost = (Q(s,a) - targetQ)^2$.
         Update Q(s, a), minimizing the cost function.
      end
      Set s = s'.
      Accumulate the reward R.
   end
   Update the target Q-network every 10 episodes.
3. Decay $\epsilon$ using $\epsilon = (\epsilon * \epsilon_{decay}, \epsilon_{end})$.
4. Save the best model based on their performance metrics

To assess the performance of the proposed TFT-DQN model, we conducted a comparative analysis with two widely used baseline algorithms: Round Robin (RR) and Weighted Round Robin (WRR). These algorithms were implemented and evaluated under the same network conditions to provide a comprehensive performance comparison. The following sections present the detailed algorithms of RR and WRR, which served as benchmarks against TFT-DQN.

Round Robin was implemented using Algorithm 2.

**Algorithm 2:** Round Robin

1. Input
      N: Number of available links.
2. Initialization
      Set current index = 0.
3. Select the link at position current index.
4. Update current index as follows:
   $current\_index = (current\_index + 1)\%N$
5. Return the selected link.

Algorithm 3 was also implemented using the Weighted Round Robin (WRR) method.

**Algorithm 3:** Weighted Round Robin

1. Input:
      W: A list of weights associated with each link, where W[i] represents the weight of link i.

2. Initialization:
      Set current index = 0.
      Set current weight = 0.

3. Increment current index as follows:
   $current\_index = (current\_index + 1)\%len(W)$

4. If current index == 0, decrement current weight by 1.

5. If current index <= 0 , set current weight to the maximum value in the weight list W.

   $current\_weight = max(W)$

6. Check if the weight of the link at current index is greater than or equal to current weight. If true, select the link at current index.
7. Repeat steps 1–4 until a link is selected.

## VII. RESULTS AND ANALYSIS

In section F under Methodology, the packet count was identified as the primary indicator of traffic load processed by the SDN controller, and thus, it was selected as the target variable for the forecasting models. Packet count represents the volume of traffic passing through the controller, and forecasting this value allows for proactive traffic management and load balancing. Additional features, such as packet size, throughput, and other flow-related statistics, were utilized as input variables, providing a more comprehensive view of the network's traffic behavior.

To optimize the dataset, Random Forest regression was employed to assess feature importance, which effectively reduced the number of features from 29 to 23, enhancing the efficiency of the analysis. Additionally, a Time index column was generated based on the data collection interval, preparing the dataset for accurate and reliable forecasting tasks. This rigorous preparation ensures the dataset's suitability for advanced machine learning applications, particularly in the context of SDN traffic load forecasting.

Figure 11 shows the distribution of key features used in the model, visualized as histograms. This helps in understanding the range and distribution of values for each feature, which is crucial for identifying potential issues such as skewness or outliers before training the model.

The heatmap in figure 12 visualizes the correlation matrix between various network metrics (such as Switch_id, In_port, Out_port, Packet_count, etc.). Correlations range from -1 to 1, with the following colour-coding:

1) *Red Shades (Positive Correlation):* Strong positive correlations are shown in deep red. These indicate that as one variable increases, the other also increases. For example, Byte_count and Packet_count have a strong positive correlation (~0.99). Bandwidth (Kbps) and both Tx_bitrate (Kbps) and Rx_bitrate(Kbps) have very high positive correlations (~0.92). Rx_avg_packet_size and Tx_avg_packet_size are highly positively correlated (~0.75).

2) *Blue Shades (Negative Correlation):* Blue shades show negative correlations. This indicates that as one variable increases, the other decreases. For instance, Packet_loss and Rx_packets are negatively correlated (~-0.53). Packet_loss also has a notable negative correlation with Rx_bytes (~-0.50). Tx_Bandwidth_Utilization and Packet_loss exhibit a weaker negative correlation (~-0.38).

3) *White or Light Shades (Near Zero Correlation):* Lighter shades (white to very pale colours) indicate weak or no significant correlation, meaning the variables are largely independent.

4) Many metrics, such as Switch_id, Flow_speed (Kbps), and Bandwidth_efficiency, show little correlation with most other metrics.

Notable Observations:
i. Traffic Metrics: Traffic-related metrics, such as Tx_packets, Rx_packets, Tx_bytes, and Rx_bytes, are generally positively correlated, which makes sense as higher transmission and reception volumes typically go hand-in-hand.
ii. Packet Loss: Packet Loss has some interesting correlations, particularly negative ones with metrics such as Rx_packets and Rx_bytes, indicating that higher packet loss is associated with fewer received packets and bytes.
iii. Bandwidth and Throughput: Bandwidth (Kbps) shows strong positive correlations with both Tx_bitrate (Kbps) and Rx_bitrate(Kbps), signifying that higher bandwidth utilization is closely tied to higher bitrates.

Overall, the color gradients help identify relationships between different network metrics, providing insights into which variables move together and which have inverse relationships.

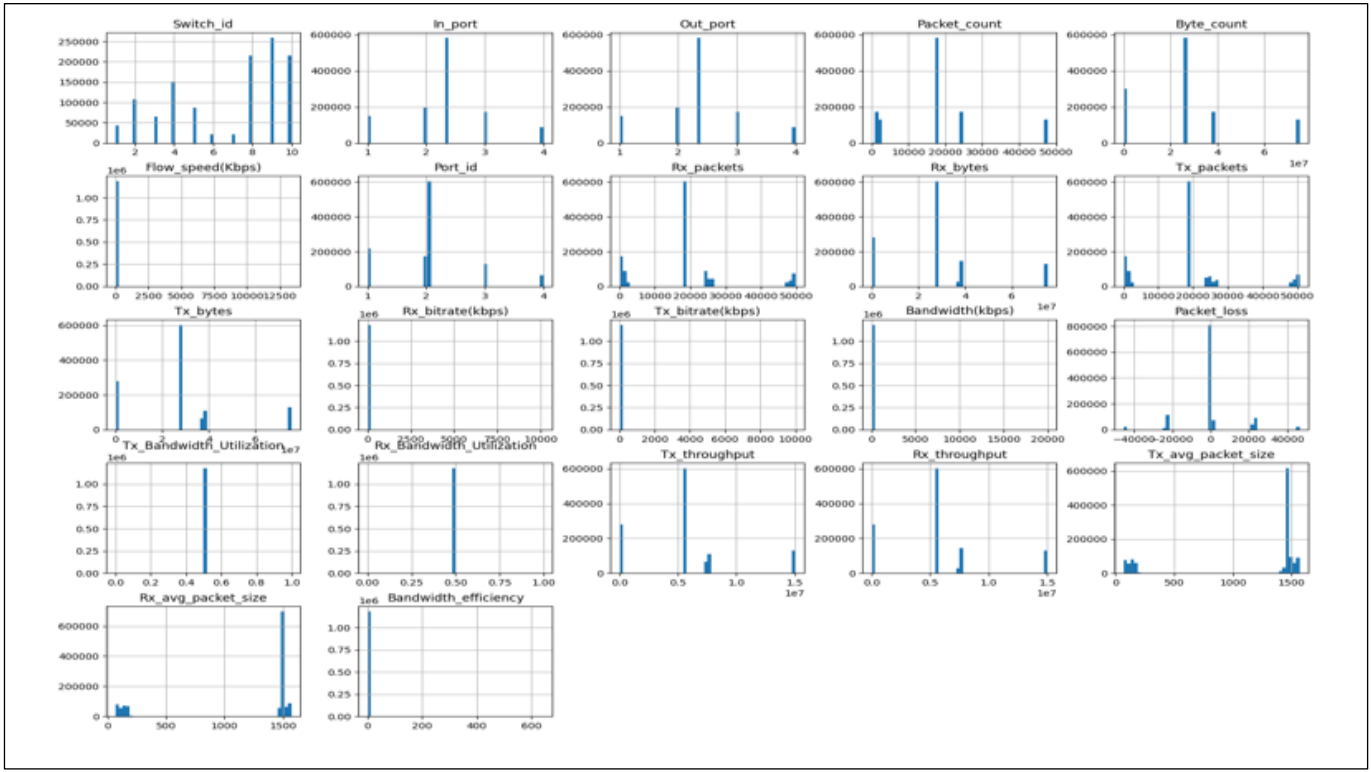

Fig. 11. Histogram of 23 features used in the model.

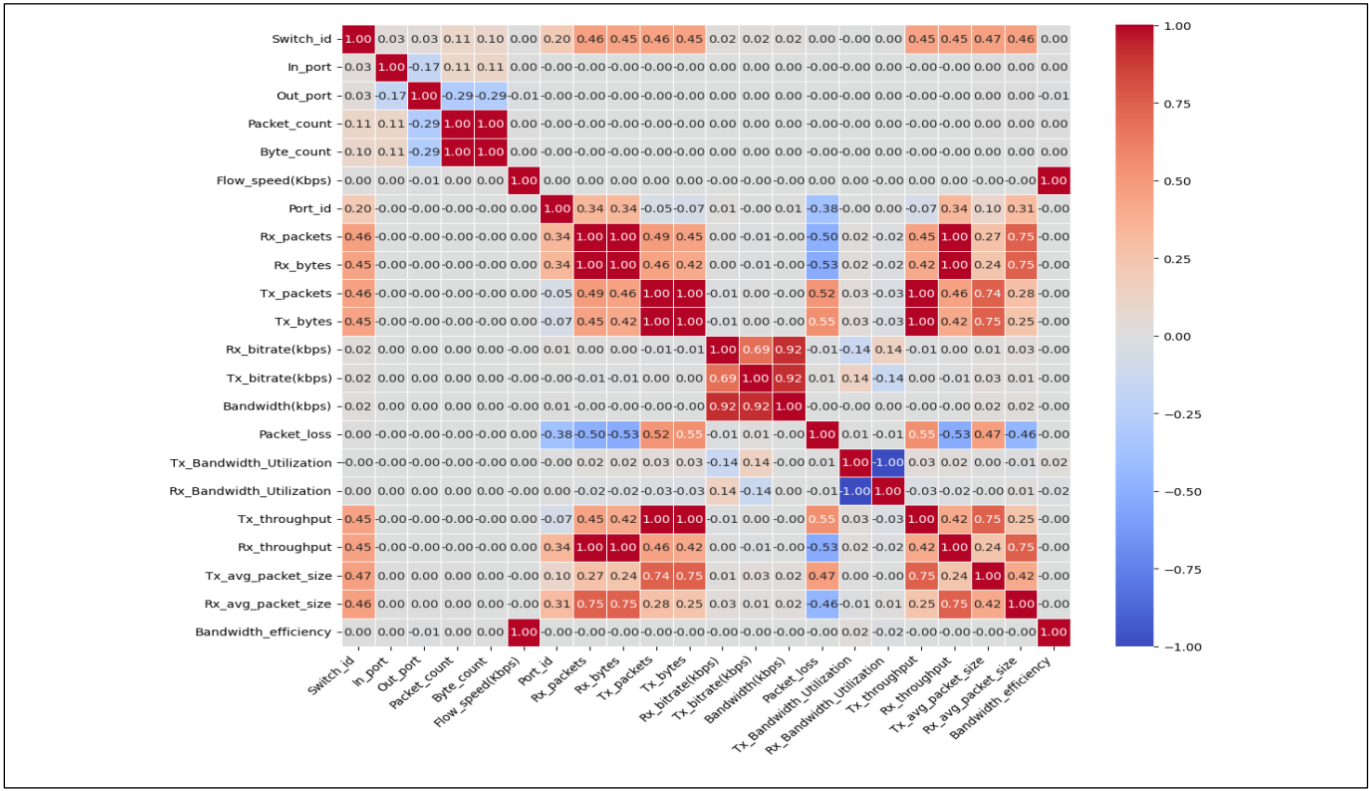

Fig. 12. Correlation matrix heatmap of the features.

After the model training process discussed in sections H and J under methodology, the performance of the Temporal Fusion Transformer (TFT) and Long Short-Term Memory (LSTM) models are evaluated first based on several key performance metrics. These metrics include Mean Absolute Error (MAE), Mean Percentage Absolute Error (MPAE), R-squared (R²) score, and Symmetric Mean Percentage Absolute Error (SMPAE). Each of these metrics offers a distinct perspective on the accuracy and efficiency of the models in forecasting traffic load in Software-Defined Networks (SDNs). The following sections explain the significance of each metric, along with their corresponding formulas.

1) *Mean Absolute Error (MAE):* The MAE measures the average magnitude of the errors between predicted and actual values, without considering their direction. It is given by the formula:

$$MAE = \frac{1}{n} \sum_{i=1}^{n} |y_i - \hat{y}_i|$$

where $y_i$ is the actual value, $\hat{y}_i$ is the predicted value, and n is the total number of observations.

2) *Mean Percentage Absolute Error (MPAE)*: MPAE calculates the percentage difference between the predicted and actual values, making it easier to interpret in terms of relative error. The formula is:

$$MPAE = \frac{1}{n} \sum_{i=1}^{n} \left( \frac{|y_i - \hat{y}_i|}{y_i} \right) \times 100$$

3) *R-squared (R²) Score:* The R² score is a statistical measure that indicates how well the predictions approximate the actual values. It is defined as:

$$R^2 = 1 - \frac{\sum_{i=1}^{n}(y_i - \hat{y}_i)^2}{\sum_{i=1}^{n}(y_i - \underline{y}_i)^2}$$

where $\underline{y}$ is the mean of the actual values.

4) *Symmetric Mean Percentage Absolute Error (SMPAE):* SMPAE provides a balanced metric by addressing the issue of asymmetric errors in MPAE. The formula is:

$$MPAE = \frac{1}{n} \sum_{i=1}^{n} \left( \frac{|y_i - \hat{y}_i|}{(y_i + \hat{y}_i) \times \frac{1}{2}} \right) \times 100$$

Each of these metrics will be used to compare the performance of TFT and LSTM models in the following sections.

A. *Comparison of TFT and LSTM Performance*

This section presents a comparison between the Temporal Fusion Transformer (TFT) and Long Short-Term Memory (LSTM) models, focusing on their performance in predicting SDN traffic load. The performance of both models was evaluated using the Mean Absolute Error (MAE), Mean Percentage Absolute Error (MPAE), Symmetric Mean Percentage Absolute Error (SMPAE), and the R-squared (R²) score. A summary of the results is provided in Table IV.

Table IV: Performance Comparison of TFT and LSTM Models

| Model | MAE | MPAE | SMPAE | R2 SCORE |
|---|---|---|---|---|
| TFT | 108.00 | 0.0402 | 0.0225 | 79.66% |
| LSTM | 705.87 | 0.387 | (Est.) | 10.41% |

As shown in Table IV, the TFT model outperformed the LSTM model across all metrics. Specifically:
a) The MAE for the TFT model was significantly lower at 108.00, compared to 705.87 for the LSTM model, indicating that TFT's predictions were much closer to the actual values.
b) Similarly, the MPAE for the TFT model was 0.0402, whereas the LSTM model exhibited a much higher error rate at 0.397, further demonstrating the higher accuracy of the TFT model in terms of percentage error.
c) For the SMPAE, the TFT model achieved 0.0225, providing a balanced measure of error, while the LSTM SMPAE was not directly calculated. However, based on its high MPAE and MAE values, the SMPAE for the LSTM model is estimated to be higher than that of the TFT model, likely around 0.35.
d) The R² score for the TFT model was 79.66%, indicating a strong fit between the predicted and actual values. In contrast, the LSTM model had a much lower R² score of 10.41%, suggesting that it struggled to capture the underlying patterns in the data.

The results clearly demonstrate that the Temporal Fusion Transformer (TFT) significantly outperforms the Long Short-Term Memory (LSTM) model in forecasting SDN traffic load. The lower error values across MAE, MPAE, and SMPAE, along with a much higher R² score, validate the effectiveness of TFT in capturing the temporal dependencies and complex patterns in the dataset. This superior performance can be attributed to the TFT model's advanced capabilities in handling multivariate time series data and leveraging both static and time-varying covariates, which is crucial in the context of SDN traffic load forecasting.

In contrast, the LSTM model, while commonly used for time-series forecasting, was not as effective in this particular task. The high error rates and low R² score indicate that LSTM was unable to generalize well to the patterns in the dataset, leading to poor predictions compared to TFT.

## B. Temporal Fusion Transformer (TFT) Prediction Analysis

Table V: Predictions made by TFT

| TIMESTEPS(s) | MAE | MPAE |
|---|---|---|
| t+1 | 0.0006036676 | 3.800061776360053e-06 |
| t+2 | 0.0006037299 | 3.800459325020711e-06 |
| t+3 | 0.0006037299 | 3.800459325020711e-06 |
| t+4 | 0.0006037299 | 3.800459325020711e-06 |
| t+5 | 0.0006037931 | 3.8008568736813686e-06 |
| t+6 | 0.0006037299 | 3.800459325020711e-06 |
| t+7 | 0.0006032248 | 3.7972800015495523e-06 |
| t+8 | 0.00060366676 | 3.800061776360053e-06 |
| t+9 | 0.0006037931 | 3.8008568736813686e-06 |
| t+10 | 0.0006037931 | 3.8008568736813686e-06 |
| t+11 | 0.0006037299 | 3.800459325020711e-06 |
| t+12 | 0.0006048298 | 3.800459325020811e-06 |

The Table V shows the predictions made by the Temporal Fusion Transformer (TFT) model across 12 timesteps, using Mean Absolute Error (MAE) and Mean Percentage Absolute Error (MPAE) as performance metrics. The maximum encoding length for the time series used in this study was 12, meaning that the model forecasts over 12 steps ahead.

Each row represents a specific timestep, ranging from t+1 to t+12. As shown, the values for MAE are relatively consistent across the timesteps, indicating that the model's predictions maintain a consistent level of accuracy over time. The MAE values remain around 0.000603, which is indicative of a relatively small absolute error in predicting the traffic load, making the model effective at capturing the underlying patterns in the data.

On the other hand, the MPAE values fluctuate slightly but stay within a small range, with the values being on the order of $10^{-6}$. This suggests that the percentage error in relation to the actual packet count remains minimal, reflecting the robustness of the TFT model in providing reliable forecasts, even at later time steps.

The values marked in red indicate the specific points where both MAE and MPAE slightly deviate from the trend. This minor deviation may indicate occasional challenges in the model's ability to maintain its accuracy. However, given the small magnitudes of the errors, these differences do not significantly impact the overall performance.

In summary, this table demonstrates that the TFT model effectively predicts traffic load over multiple time steps, maintaining stable performance across the forecasting horizon. The low error metrics reinforce the suitability of the model for time series forecasting tasks in the context of SDN traffic load.

## C. Model Interpretability Of The Temporal Fusion Transformer (TFT)

Model interpretability is a crucial aspect of understanding how machine learning models make predictions and their decision-making processes. For the Temporal Fusion Transformer (TFT) model, interpretability involves analyzing how the model's predictions align with actual data and assessing the relationships between input features and forecasted values.

In this study, the TFT model's interpretability is explored through visual comparisons of predicted versus actual values of key input features. The following images illustrate these comparisons, providing insights into how well the TFT model captures the underlying patterns and trends in the data.

1) *Actual vs. Predicted Values:* The provided images show the actual versus predicted values for selected input features over various timesteps. By examining these plots, we can evaluate the accuracy of the TFT model in forecasting and how closely its predictions match the real data. The consistency between the actual and predicted values indicates the model's effectiveness in capturing the temporal dynamics of the input features.

2) *Feature Analysis:* The plots further enable an analysis of the impact of different features on the model's predictions. By visualizing how changes in input features correlate with predicted values, we gain a better understanding of the TFT model's behavior and its sensitivity to different input variables.

These visualizations are essential for validating the model's performance and ensuring that it provides reliable and interpretable predictions. They help in identifying any potential discrepancies or areas where the model might need further refinement. Additionally, they support the overall analysis by highlighting the TFT model's capability to handle complex time series data and its effectiveness in predicting network traffic load.

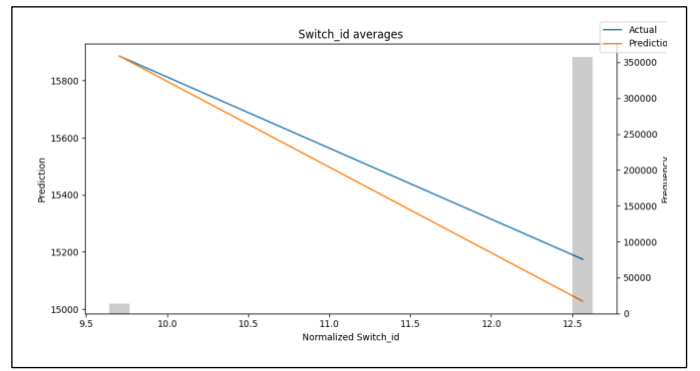

Fig. 13.  Switch id importance

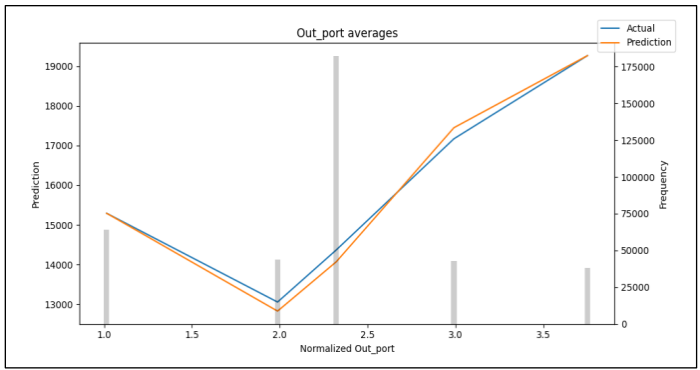

Fig. 14. Out port importance

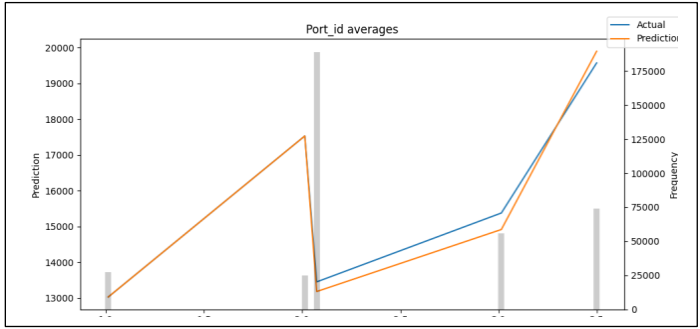

Fig. 15. Port id importance

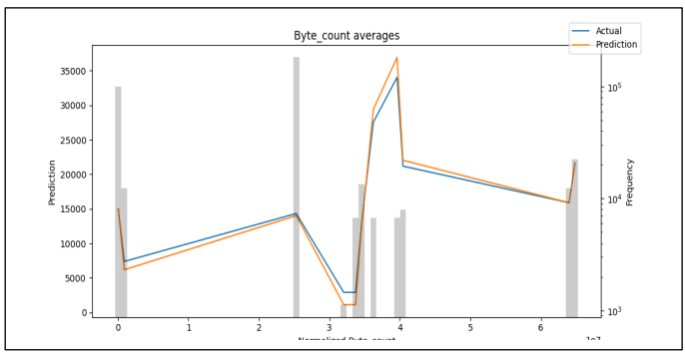

Fig. 16. Byte count importance

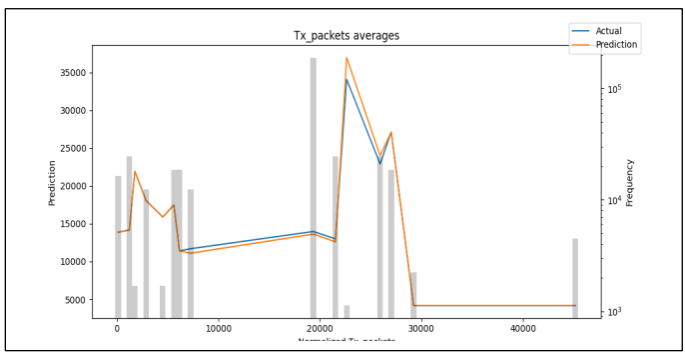

Fig. 17. Transmit packets importance

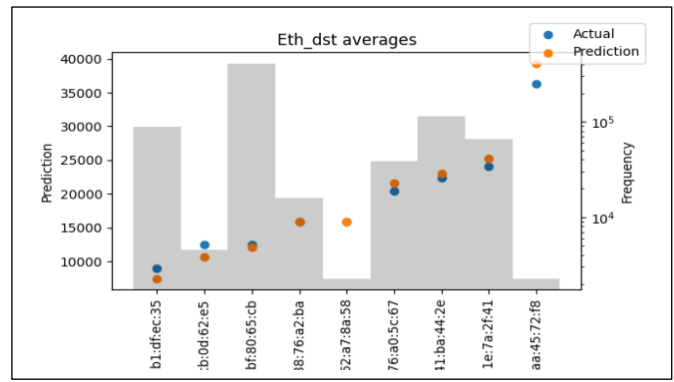

Fig. 18. Ethernet destination address importance

Figure 13 - 18 demonstrate that the TFT model maintains a robust level of accuracy across different input features and timesteps. This visual evidence reinforces the quantitative metrics (such as MAE and MPAE) and provides a comprehensive view of the model's interpretability.

The TFT model's ability to provide accurate and interpretable predictions is crucial for its application in forecasting tasks. The visualizations of actual versus predicted values enhance our understanding of the model's performance and its alignment with the underlying data patterns.

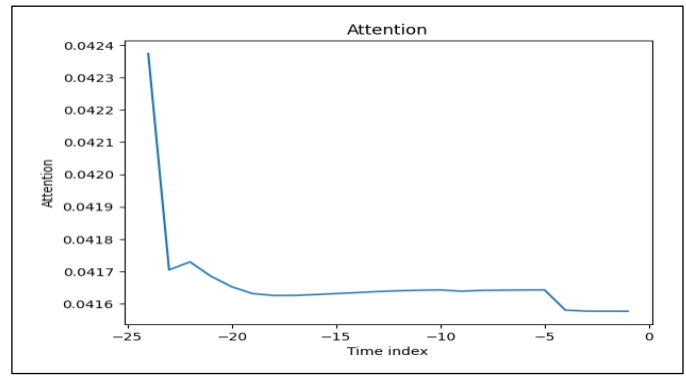

Fig. 19. Attention Graph

The attention graph in figure 19 illustrates the distribution of attention weights over the time index during the prediction process.

The y-axis represents the attention values, while the x-axis shows the time index, with negative values representing time steps before the current forecast. Higher attention weights indicate that the model places more emphasis on those particular time steps for making accurate predictions.

From figure 19, we observe that the model initially places significant attention on the earlier time steps, with a sharp decline afterward. This suggests that the earlier time indices contribute more to the model's decision-making process, likely because they contain more valuable information for predicting future states. As time progresses, the attention becomes more stable and slightly decreases, indicating that recent data is less influential compared to earlier information in this specific instance.

This pattern provides insight into how the TFT model utilizes historical data to make predictions, assigning different importance to various time points to enhance its performance.

*D. Feature Importance Analysis*

The graphs in figures 20, 21 and 22 respectively represent static, encoder and decoder variable importance which give crucial insights into which variables the model relied on most during the training and forecasting process. Here's a brief analysis of the graphs:

1) *Static Variable Importance*
   a) The most significant static variable in this context is Eth_dst (Ethernet destination address), which accounts for over 30% of the importance. This suggests that the destination of Ethernet frames is a crucial factor for traffic forecasting.
   b) Link_id, In_port, and Out_port also contribute significantly, indicating the importance of link identifiers and port information for directing traffic within the SDN setup.

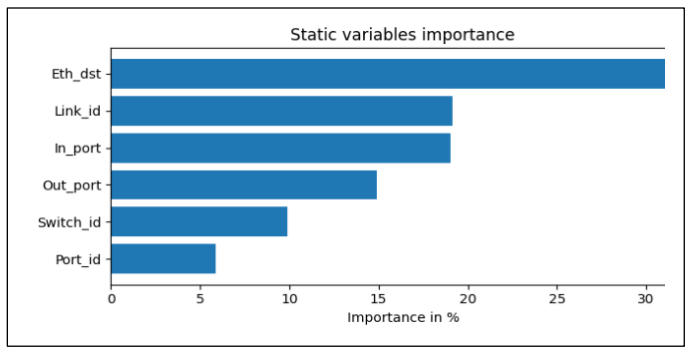

Fig. 20. Static Variable Importance

2) Encoder Variable Importance
   a) Among the encoder variables, Rx_Bandwidth_Utilization (received bandwidth utilization) has the highest importance, implying that the utilization of received bandwidth plays a major role in determining traffic flow and congestion in the network.
   b) Tx_packets and Tx_bitrate(kbps) follow closely, indicating that both transmitted packet counts and their respective bitrate are key factors in forecasting network behavior.
   c) Variables like Packet_loss, Rx_bitrate(kbps), and Tx_avg_packet_size are also relevant, reflecting the model's sensitivity to network quality metrics such as loss and bitrate.

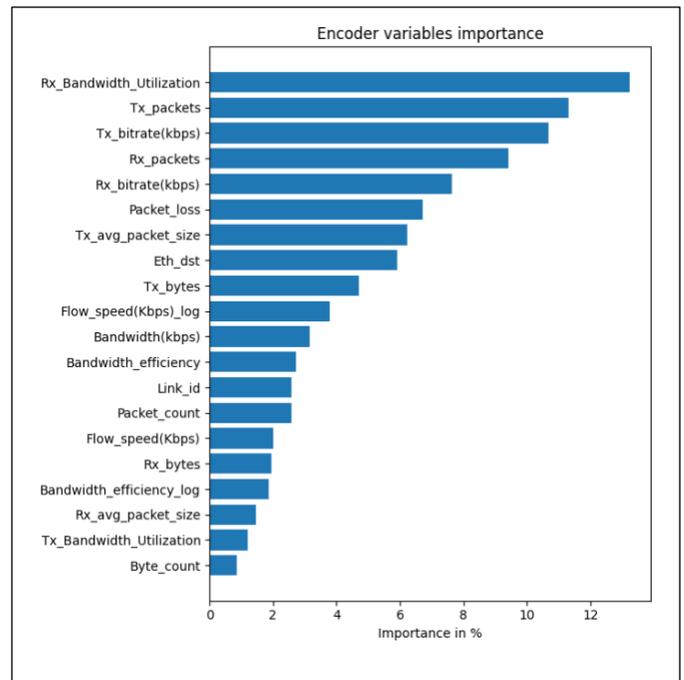

Fig. 21. Encoder Variable Importance

3) *Decoder Variable Importance*
   a) The analysis of the decoder variables highlights which features contribute most to the model's predictive power in the Temporal Fusion Transformer (TFT) architecture. The graph shows that Rx_Bandwidth_Utilization holds the highest importance, with an influence of nearly 14%. This is consistent with expectations, as the model must prioritize understanding the incoming bandwidth utilization to accurately predict traffic loads.
   b) The second most important feature is Tx_bitrate(Kbps), which closely follows at around 13%. This indicates that the transmission rate plays a key role in traffic prediction, reflecting how fast data is transmitted over the network.
   c) Next, Tx_packets and Tx_avg_packet_size are the subsequent significant contributors, both exceeding 10% in importance. These features encapsulate information about the size and number of packets being transmitted, which are critical for load balancing and anticipating network congestion. The high placement of Rx_packets further solidifies the importance of packet-level data in determining the traffic patterns.
   d) Other features such as Bandwidth_efficiency_log, Tx_bytes, Bandwidth (Kbps), and Packet_loss also carry substantial weight, each contributing to the model's predictions. Packet_loss, in particular, offers insight into the network's reliability, while Tx_bytes and Bandwidth_efficiency provide additional metrics of data transmission quality and usage.

Interestingly, variables such as Flow_speed(Kbps)_log, Link_id, and Byte_count also feature in the list, though with lower relative importance. These metrics still contribute to the model's understanding but are likely supplementary to the more prominent bandwidth and packet-related features.

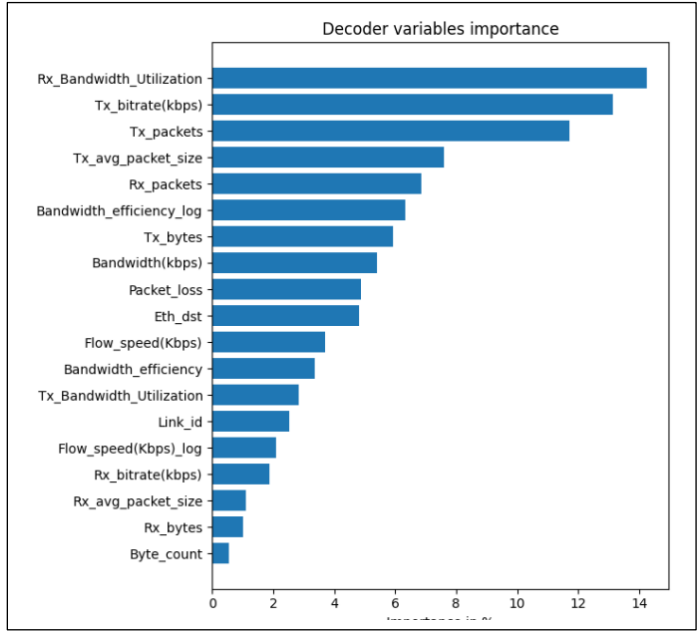

Fig. 22. Decoder Variable Importance

These analyses provide an understanding of which features the TFT model emphasizes, allowing us to fine-tune network management decisions based on the most impactful factors in SDN load balancing and traffic forecasting.

*E. DQN Model Results and Analysis*

As mentioned in the section L of the methodology where the training of the DQN model was carried out, this section presents the performance of the Deep Q-Network (DQN) model and it is evaluated based on several key performance metrics, including reward function analysis, throughput, packet loss, and latency during training. Additionally, we compare the DQN model against Round Robin (RR) and Weighted Round Robin (WRR) scheduling algorithms in two different simulation environments (500MB and 1000MB). The results highlight the effectiveness of the DQN model in improving network performance for dynamic load balancing.

1) *Training Analysis*

During the DQN training phase, several performance metrics were monitored, including reward, throughput, packet loss, and latency. The reward function provides insights into how well the agent is learning over time, aiming to maximize throughput while minimizing both latency and packet loss.

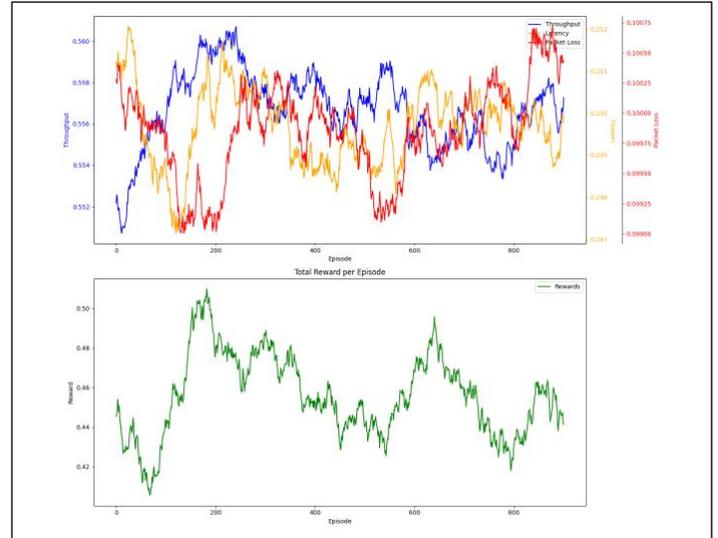

Fig. 23. Training Graphs

a) *Reward Function Analysis:* Figure 23 shows that the best performance occurs between episode 200 and 600, where the agent successfully maximized throughput and minimized latency and packet loss. This is evident from the corresponding throughput, latency, and packet loss graphs in this episode range. The throughput was highest, while latency and packet loss were at their lowest values during this interval, indicating optimal performance by the DQN agent.

b) *Metrics during Episode 200-600*: The interval between episode 200 and 600 is shown in Figure 24, highlighting the metrics' behaviour:

  i. Higher throughput values during training indicate better network utilization.

  ii. Lower latency values show the agent's ability to maintain low delays.

  iii. Low packet loss demonstrates the effectiveness of the agent in minimizing dropped packets.

c) *Action Distribution:* Another key aspect of the DQN model is the action distribution graph, which illustrates the various links selected by the agent during training (Figure 25). This distribution reflects how the agent dynamically selects the optimal links to maximize the reward, balancing load across the network.

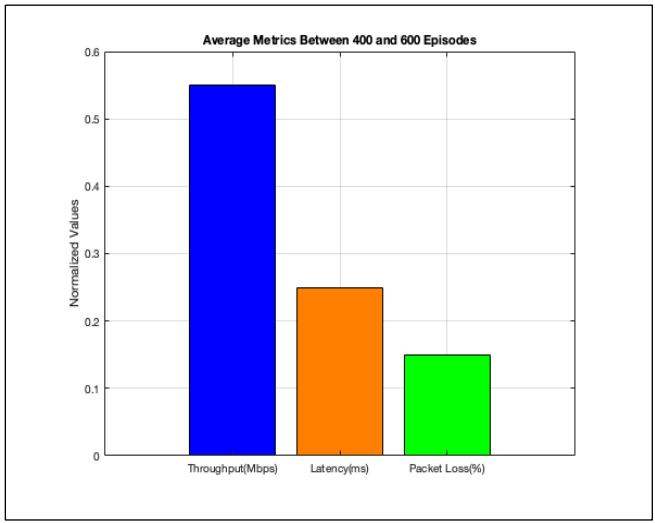

Fig. 24.  Measured QoS

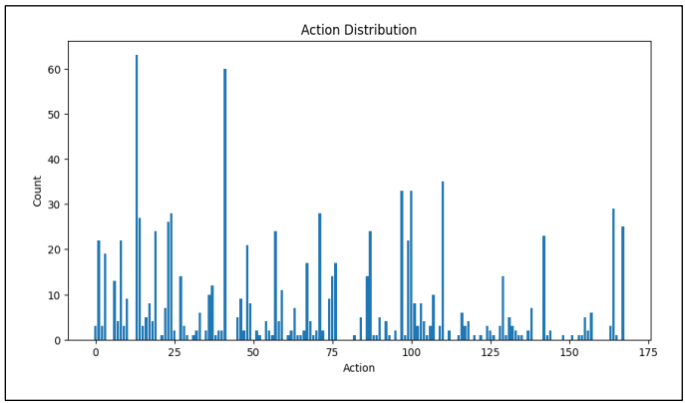

Fig. 25.  Action Distribution

2) *Prediction Performance Analysis:* The TFT-DQN model's performance was compared to RR and WRR algorithms across two different simulation environments with data rates of 500MB and 1000MB. The comparison was based on normalized values of throughput, latency, and packet loss.

   a) Performance on 500MB Data Rate Simulation

   i. Throughput: The average throughput achieved by the DQN model was 0.275, outperforming RR (0.202) and WRR (0.205). This demonstrates the DQN's ability to efficiently utilize network resources and manage load more effectively than traditional algorithms.

   ii. Latency: The DQN model showed a significant reduction in latency, with an average latency of 0.105, compared to 0.176 for RR and 0.181 for WRR. This improvement suggests that the DQN model optimizes packet routing to minimize delays in the network.

   iii. Packet Loss: The DQN model achieved the lowest packet loss at 0.080, compared to RR (0.174) and WRR (0.121). The reduction in packet loss further highlights the DQN's ability to handle traffic more efficiently, resulting in fewer dropped packets.

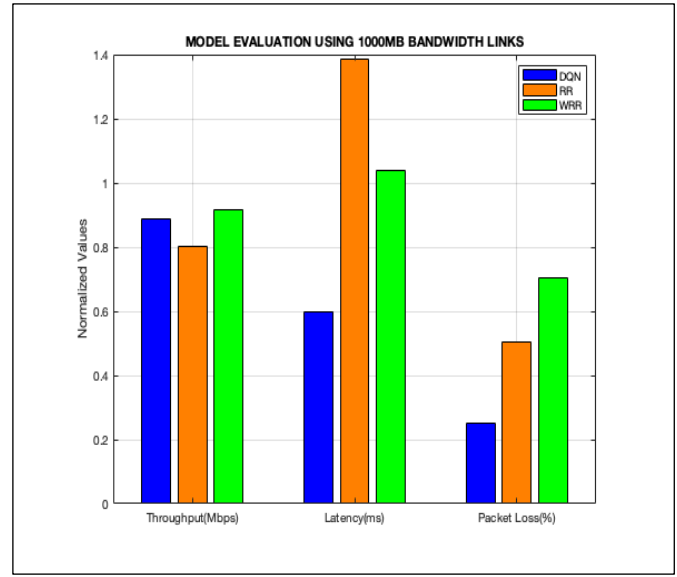

Fig. 26. Model Evaluation using 500MB Bandwidth

   b) Performance on 1000MB Data Rate Simulation

   i. Throughput: In the 1000MB environment, the DQN model achieved an average throughput of 0.885, higher than RR (0.802) and comparable to WRR (0.918). This indicates that DQN can handle larger data rates efficiently, ensuring high throughput similar to WRR.

   ii. Latency: The DQN model again outperformed both RR and WRR in terms of latency, with an average latency of 0.600, compared to 1.384 for RR and 1.04 for WRR. This demonstrates the DQN's superior ability to reduce delays in the network under higher data rates.

   iii. Packet Loss: The average packet loss for the DQN model was 0.250, significantly lower than RR (0.506) and WRR (0.706). This lower packet loss highlights the model's enhanced performance in preventing packet drops even in high data rate environments.

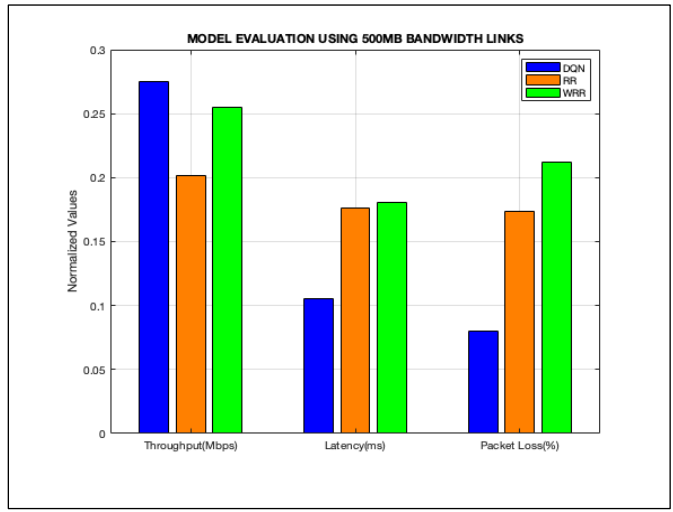

Fig. 27. Model Evaluation using 1000MB Bandwidth

From the results, it is clear that the DQN model outperforms both the RR and WRR scheduling algorithms across key metrics such as throughput, latency, and packet loss in both the 500MB and 1000MB simulation environments (shown in figures 26 and 27). The TFT-DQN approach provides a more dynamic and efficient solution for load balancing in Software-Defined Networks (SDNs), adapting to changing network conditions and optimizing performance.

## VIII. CONCLUSION

In this study, we addressed the challenge of load balancing in Software-Defined Networks (SDNs) by utilizing advanced machine learning techniques, specifically the Temporal Fusion Transformer (TFT) for traffic forecasting and the Deep Q-Network (DQN) for dynamic load balancing. The findings demonstrated that traditional load balancing methods, such as Round Robin (RR) and Weighted Round Robin (WRR), lack the flexibility required to handle dynamic and complex network environments. In contrast, SDNs, with their separation of the control and data planes, offer a more adaptive solution by enabling real-time traffic analysis and optimization. The integration of the TFT and DQN models resulted in significant improvements in network performance, including higher throughput, reduced latency, and minimized packet loss. The TFT model predicted future network traffic patterns, while the DQN dynamically adjusted routing decisions, ensuring optimal load distribution. The TFT-DQN model consistently outperformed traditional algorithms, showcasing the potential of machine learning to revolutionize load balancing in modern networks. Future research should explore expanding the model's generalization across various network topologies, incorporating hybrid machine learning approaches, and optimizing energy efficiency to further enhance the applicability of this model in large-scale SDN deployments. This study contributes to the growing field of intelligent network management, providing a foundation for more adaptive and efficient traffic engineering solutions in next-generation networks.

REFERENCES

[1] D. Kreutz, F. M. V. Ramos, P. E. Verissimo, C. E. Rothenberg, S. Azodolmolky, and S. Uhlig, "Software-defined networking: A comprehensive survey," *Proceedings of the IEEE*, vol. 103, no. 1, pp. 14–76, 2015, doi: 10.1109/JPROC.2014.2371999.

[2] F. Hu, Q. Hao, and K. Bao, "A survey on software-defined network and OpenFlow: From concept to implementation," *IEEE Communications Surveys and Tutorials*, vol. 16, no. 4, pp. 2181–2206, 2014, doi: 10.1109/COMST.2014.2326417.

[3] R. K. Das, F. H. Pohrmen, A. K. Maji, and G. Saha, "FT-SDN: A Fault-Tolerant Distributed Architecture for Software Defined Network," *Wirel Pers Commun*, vol. 114, no. 2, pp. 1045–1066, 2020, doi: 10.1007/s11277-020-07407-x.

[4] J. N. Iyanda, "Analysis and Evaluation of Quality of Service ( QoS ) Router using Round Robin ( RR ) and Weighted Round Robin ( WRR )," vol. 5, no. 2, pp. 1–9, 2015.

[5] S. B. Vyakaranal and J. G. Naragund, "Weighted Round-Robin Load Balancing Algorithm for Software-Defined Network BT - Emerging Research in Electronics, Computer Science and Technology," V. Sridhar, M. C. Padma, and K. A. R. Rao, Eds., Singapore: Springer Singapore, 2019, pp. 375–387.

[6] R. Kumar, V. U., and V. Tiwari, "Optimized traffic engineering in Software Defined Wireless Network based IoT (SDWN-IoT): State-of-the-art, research opportunities and challenges," *Comput Sci Rev*, vol. 49, p. 100572, Aug. 2023, doi: 10.1016/J.COSREV.2023.100572.

[7] "Artificial Intelligence based Load balancing in SDN: A Comprehensive Survey."

[8] B. Lim, S. Arık, N. Loeff, and T. Pfister, "Temporal Fusion Transformers for interpretable multi-horizon time series forecasting," *Int J Forecast*, vol. 37, no. 4, pp. 1748–1764, Oct. 2021, doi: 10.1016/j.ijforecast.2021.03.012.

[9] M. R. Belgaum, S. Musa, M. M. Alam, and M. M. Su'Ud, "A Systematic Review of Load Balancing Techniques in Software-Defined Networking," *IEEE Access*, vol. 8, pp. 98612–98636, 2020, doi: 10.1109/ACCESS.2020.2995849.

[10] O. G. Matlou and A. M. Abu-Mahfouz, "Utilising artificial intelligence in software defined wireless sensor network," in *IECON 2017 - 43rd Annual Conference of the IEEE Industrial Electronics Society*, 2017, pp. 6131–6136. doi: 10.1109/IECON.2017.8217065.

[11] S. Almakdi, A. Aqdus, R. Amin, and M. S. Alshehri, "An Intelligent Load Balancing Technique for Software Defined Networking Based 5G Using Machine Learning Models," *IEEE Access*, vol. 11, no. August, pp. 105082–105104, 2023, doi: 10.1109/ACCESS.2023.3317513.

[12] S. Bhardwaj and A. Girdhar, "Software-Defined Networking: A Traffic Engineering Approach," in *2021 IEEE 8th Uttar Pradesh Section International Conference on Electrical, Electronics and Computer Engineering (UPCON)*, 2021, pp. 1–5. doi: 10.1109/UPCON52273.2021.9667584.

[13] A. Mohammed *et al.*, "Weighted Round Robin Scheduling Algorithms in Mobile AD HOC Network," *HORA 2021 - 3rd International Congress on Human-Computer Interaction, Optimization and Robotic Applications, Proceedings*, pp. 3–7, 2021, doi: 10.1109/HORA52670.2021.9461358.

[14] V. Tosounidis, G. Pavlidis, and I. Sakellariou, "Deep Q-learning for load balancing traffic in SDN networks," *ACM International Conference Proceeding Series*, pp. 135–143, 2020, doi: 10.1145/3411408.3411423.

[15] B. Haidi, T. W. Au, and S. H. Newaz, "Single Cluster Load Balancing Using SDN: Performance Comparison between Floodlight and POX," 2019, pp. 1332–1336. doi: 10.1109/ICCT46805.2019.8947274.

[16] I. F. Akyildiz, A. Lee, P. Wang, M. Luo, and W. Chou, "A roadmap for traffic engineering in SDN-OpenFlow networks," *Computer Networks*, vol. 71, pp. 1–30, Oct. 2014, doi: 10.1016/J.COMNET.2014.06.002.

[17] M. Karakus and A. Durresi, "Quality of Service (QoS) in Software Defined Networking (SDN): A survey," *Journal of Network and Computer Applications*, vol. 80, pp. 200–218, Feb. 2017, doi: 10.1016/J.JNCA.2016.12.019.

[18] C. X. Cui and Y. Bin Xu, "Research on load balance method in SDN," *International Journal of Grid and Distributed Computing*, vol. 9, no. 1, pp. 25–36, 2016, doi: 10.14257/ijgdc.2016.9.1.03.

[19] L. Davoli, L. Veltri, P. L. Ventre, G. Siracusano, and S. Salsano, "Traffic Engineering with Segment Routing: SDN-Based Architectural Design and Open Source Implementation," in *2015 Fourth European Workshop on Software Defined Networks*, 2015, pp. 111–112. doi: 10.1109/EWSDN.2015.73.

[20] O. Dobrijevic, M. Santl, and M. Matijasevic, "Ant colony optimization for QoE-centric flow routing in software-defined networks," in *2015 11th International Conference on Network and Service Management (CNSM)*, 2015, pp. 274–278. doi: 10.1109/CNSM.2015.7367371.

[21] P. Wang, S.-C. Lin, and M. Luo, "A Framework for QoS-aware Traffic Classification Using Semi-supervised Machine Learning in SDNs," in *2016 IEEE International Conference on Services Computing (SCC)*, 2016, pp. 760–765. doi: 10.1109/SCC.2016.133.

[22] W. Xiang, N. Wang, and Y. Zhou, "An Energy-Efficient Routing Algorithm for Software-Defined Wireless Sensor Networks," *IEEE Sens J*, vol. 16, no. 20, pp. 7393–7400, 2016, doi: 10.1109/JSEN.2016.2585019.

[23] V. Srivastava and R. S. Pandey, "Machine intelligence approach: To solve load balancing problem with high quality of service performance for multi-controller based Software Defined Network," *Sustainable Computing: Informatics and Systems*, vol. 30, p. 100511, Jun. 2021, doi: 10.1016/J.SUSCOM.2021.100511.

[24] Z. Li, X. Zhou, J. Gao, and Y. Qin, "SDN Controller Load Balancing Based on Reinforcement Learning," *Proceedings of the IEEE International Conference on Software Engineering and Service Sciences, ICSESS*, vol. 2018-Novem, pp. 1120–1126, 2018, doi: 10.1109/ICSESS.2018.8663757.

[25] A. Filali, Z. Mlika, S. Cherkaoui, and A. Kobbane, "Preemptive SDN Load Balancing with Machine Learning for Delay Sensitive Applications," *IEEE Trans Veh Technol*, vol. 69, no. 12, pp. 15947–15963, 2020, doi: 10.1109/TVT.2020.3038918.

[26] A. Zeng, M. Chen, L. Zhang, and Q. Xu, "Are Transformers Effective for Time Series Forecasting?," *Proceedings of the 37th AAAI Conference on Artificial Intelligence, AAAI 2023*, vol. 37, pp. 11121–11128, 2023, doi: 10.1609/aaai.v37i9.26317.

[27] Q. Wen *et al.*, "Transformers in Time Series : A Survey," 2022.

[28] D. Bermúdez Garzón, C. Gómez, M. E. Gómez, P. López, and J. Duato, "LNCS 7484 - Towards an Efficient Fat–Tree like Topology," 2012.

[29] H. Mahmood-Khan, "Fat-Tree Data Center Topology using Mininet." [Online]. Available: https://github.com/HassanMahmoodKhan/Fat-Tree-Data-Center-Topology?tab=readme-ov-file#start-of-content

[30] L. Cai, K. Janowicz, G. Mai, B. Yan, and R. Zhu, "Traffic transformer: Capturing the continuity and periodicity of time series for traffic forecasting," *Transactions in GIS*, vol. 24, no. 3, pp. 736–755, Jun. 2020, doi: 10.1111/tgis.12644.

[31] A. R. Abdellah, O. A. K. Mahmood, A. Paramonov, and A. Koucheryavy, "IoT traffic prediction using multi-step ahead prediction with neural network," in *2019 11th International Congress on Ultra Modern Telecommunications and Control Systems and Workshops (ICUMT)*, 2019, pp. 1–4. doi: 10.1109/ICUMT48472.2019.8970675.

[32] A. Bayati, K. Khoa Nguyen, and M. Cheriet, "Multiple-Step-Ahead Traffic Prediction in High-Speed Networks," *IEEE Communications Letters*, vol. 22, no. 12, pp. 2447–2450, 2018, doi: 10.1109/LCOMM.2018.2875747.

[33] Y. Yu, X. Si, C. Hu, and J. Zhang, "A Review of Recurrent Neural Networks: LSTM Cells and Network Architectures," *Neural Comput*, vol. 31, no. 7, pp. 1235–1270, Jul. 2019, doi: 10.1162/neco_a_01199.

[34] Y. Li, "DEEP REINFORCEMENT LEARNING: AN OVERVIEW," 2018. [Online]. Available: https://arxiv.org/abs/

[35] A. A. Ali, "An Introduction to Q-Learning: A Tutorial For Beginners." [Online]. Available: https://www.datacamp.com/tutorial/introduction-q-learning-beginner-tutorial

[36] U. Siddiqi, S. Sait, and M. Uysal, "Deep Reinforcement Based Power Allocation for the Max-Min Optimization in Non-Orthogonal Multiple Access," 2020.